\documentclass[multphys,vecphys]{svmult}

\usepackage{makeidx}     
\usepackage{graphicx}    
\usepackage{multicol}    

\makeindex             

\begin{document}

\title*{Understanding Type II Supernovae}
\author{L. Zampieri\inst{1}, M. Ramina\inst{1,2} and A. Pastorello\inst{1,2}}
\authorrunning{Understanding Type II Supernovae}
\institute{INAF-Osservatorio Astronomico di Padova, Vicolo
dell'Osservatorio 5, I-35122 Padova, Italy
\and Dipartimento di Astronomia, Universit\`a di Padova,  Vicolo
dell'Osservatorio 2, I-35122 Padova, Italy}
%
%
\maketitle

We present the results of a systematic analysis of a group of Type II
plateau supernovae that span a large range in luminosities, from faint
objects like SN 1997D and 1999br to very luminous events like SN
1992am. The physical properties of the supernovae appear to be related
to the plateau luminosity or the expansion velocity. The simultaneous
analysis of the observed light curves, line velocities and continuum
temperatures leads us to robust estimates of the physical parameters
of the ejected envelope. We find strong correlations among several
parameters. The implications of these results regarding the nature of
the progenitor, the central remnant and the Ni yield are also
addressed.

\section{Introduction}
\label{sec:1}

Type II supernovae (SNe) are believed to be core-collapse SNe
originating from massive ($> 8 M_\odot$) red supergiants that retain
their Hydrogen (H) envelopes. The overall phenomenological appearance
of these SNe is rather well understood (see e.g. \cite{arn96}).
However, despite lightcurve and spectral modelling have provided
important information on the physical properties of single objects
(see e.g \cite{woosley88}), comparatively little effort has been
devoted to study the correlations between the basic properties of Type
II SNe and to understand to what extent the variety of their
observational properties can be explained in terms of continuous
changes of some fundamental physical variables.
This is especially interesting after the recent discovery of a group
of low luminosity (LL), $^{56}$Ni poor SNe \cite{past03,zamp03}, whose
relation with the ``normal'' and more luminous Type II events is still
under debate. The work in this area has certainly been hampered also
by the very heterogeneous behavior of Type II SNe. However, a recent
investigation has shown that significant correlations exist among the
plateau luminosity, the expansion velocity measured at 50 days after
the explosion and the ejected $^{56}$Ni mass \cite{hamuy03}. Here we
present the results of a systematic analysis of a group of Type II
plateau supernovae that extends, especially at very low luminosity,
the sample previously considered. While we confirm
the results of Hamuy \cite{hamuy03}, we do not find evidence of a
definite correlation between the ejected envelope mass and the other
parameters.

\section{Selected Sample of Type II Plateau SNe}
\label{sec:2}

The data were taken from literature and/or extracted from the large
database of lightcurves and spectra of the Padova-Asiago Supernova
Archive. A description of the selection process is outlined in
Pastorello et al. (these Proceedings). Observations of SN
2003Z\footnote{Made in part at the Telescopio Nazionale Galileo (TNG)
under program TAC\_48.}, the first LL event extensively monitored from
explosion up to the nebular stage, are also included in this work. SNe
with uncertain estimates of the distance and interstellar absorption
and/or with signs of significant interaction with the circumstellar
material were not considered. The main selection criterion was to
choose objects that cover a big range in luminosity, including LL,
97D-like events \cite{past03,zamp03} and luminous 92am-like objects
\cite{schmidt94}. The selected objects are reported in
Table~\ref{tab:2}. Most of them have a good photometric coverage until
300--400 days after the explosion and at least 4--5 spectra in the
photosperic phase (up to $\sim 100-120$ days). The best available
estimates of the explosion epoch, the distance modulus and
interstellar absorption (Galactic and internal) for these objects are
reported in Pastorello et al. (these Proceedings) and Ramina (Laurea
Thesis, unpublished). SN 1987A is included for comparison.

%
%

\section{Modelling Core-collapse SNe}
\label{sec:3}

In the present analysis the physical parameters of the selected sample
of SNe are derived comparing the observational data to model
calculations. The adopted model is a semi-analytic code that solves
the energy balance equation for a spherically symmetric, homologously
expanding envelope at constant density \cite{zamp03}.
The initial conditions are rather idealized and provide an approximate
description of the ejected material after shock (and possible reverse
shock) passage, as derived from hydrodynamical calculations. In
particular, elements are assumed to be completely mixed throughout the
envelope and their distribution depends only on the coordinate
mass. Hydrogen, Helium, Carbon and Oxygen are assumed to be uniformly
distributed, whereas $^{56}$Ni is more centrally peaked. The evolution
of the expanding envelope is computed including all the relevant
energy sources powering the SN and is schematically divided in 3
phases from the photospheric up to the late nebular stages (for more
details see Zampieri et al. \cite{zamp03} and Ramina [Laurea Thesis,
unpublished]). The most important quantities computed by the code are
the light curve and the evolution of the line velocity and continuum
temperature at the photosphere. The physical properties of the
envelope are derived by performing a simultaneous fit of these three
observables with model calculations.

\section{Correlations Among Physical Parameters}
\label{sec:4}

The physical parameters of the post-shock, ejected envelope are listed
in Table~\ref{tab:2}. Only some of them are input parameters ($R_0$,
$M_{env}$, $V_0$, $T_{eff}$), while the others are computed by the
code or fixed by the observations. Two other input physical constants
are the fraction of the initial energy that goes into kinetic energy
$f_0$ and the gas opacity $\kappa$. In this calculation we adopt
$f_0=0.5$ (initial equipartition between thermal and kinetic energies)
and $\kappa=0.2$ cm$^2$ g$^{-1}$ (appropriate for an envelope
comprised of He and iron-group elements). The color correction factor
$f_c=T_c/T_{eff}$, that measures the deviation of the continuum
radiation temperature $T_c$ from the blackbody effective temperature
$T_{eff}$, was kept fixed and equal to $1.2$.

\begin{table}
\centering
\caption{Physical parameters from the semi-analytic model}
\label{tab:2}
\begin{tabular}{ccccccccc}
\hline\noalign{\smallskip}
 & $R_0$ & $M_{env}$ & $M_{Ni}$ & $V_0$ & $E$ &
 $t_{rec,0}$ & $T_{eff}$ & $\log L_p$ \\
 & ($10^{12}$ cm) & ($M_\odot$) &
 ($M_\odot$) & ($10^8$ cm s$^{-1}$) & ($10^{51}$ erg) & (days) & (K) & \\
\noalign{\smallskip}\hline\noalign{\smallskip}
1992am &$41\;^{+6}_{-5}$&$26\;^{+8}_{-3}$&$0.41\;^{+0.04}_{-0.04}$&$5.1\;^{+0.5}_{-0.4}$&$8.1\;^{+4.1}_{-2.0}$           & 53 &$4400\;^{+400}_{-300}$ & 42.55 \\
1992H &$38\;^{+3}_{-2}$&$23\;^{+7}_{-3}$&$0.18\;^{+0.01}_{-0.01}$&$4.9\;^{+0.2}_{-0.4}$&$6.6\;^{+2.7}_{-1.8}$           & 50 &$4300\;^{+200}_{-200}$ & 42.4 \\
1996W &$37\;^{+5}_{-3}$&$16\;^{+4}_{-2}$&$0.17\;^{+0.02}_{-0.02}$&$4.1\;^{+0.3}_{-0.3}$&$3.2\;^{+1.3}_{-0.8}$           & 48 &$4500\;^{+400}_{-300}$ & 42.25 \\
1995ad &$17\;^{+3}_{-2}$&$12\;^{+2}_{-2}$&$0.029\;^{+0.003}_{-0.004}$&$4.0\;^{+0.4}_{-0.4}$&$2.3\;^{+0.9}_{-0.7}$       & 30 &$4700\;^{+300}_{-200}$ & 41.9 \\
1969L &$25\;^{+3}_{-2}$&$16\;^{+2}_{-1}$&$0.067\;^{+0.006}_{-0.005}$&$3.6\;^{+0.2}_{-0.2}$&$2.5\;^{+0.6}_{-0.4}$        & 50 &$4300\;^{+200}_{-10}$  & 42.0 \\
1987A &$6\;^{+0.9}_{-0.7}$&$18\;^{+4}_{-2}$&$0.075\;^{+0.006}_{-0.006}$&$2.8\;^{+0.2}_{-0.2}$&$1.7\;^{+0.6}_{-0.4}$     & 26 &$4300\;^{+100}_{-200}$ & 41.35 \\
1996an &$19\;^{+2}_{-3}$&$13\;^{+2}_{-1}$&$0.050\;^{+0.005}_{-0.005}$&$3.3\;^{+0.1}_{-0.2}$&$1.7\;^{+0.3}_{-0.3}$       & 46 &$4200\;^{+200}_{-100}$ & 41.8 \\
1999em &$14\;^{+3}_{-2}$&$14\;^{+2}_{-1}$&$0.022\;^{+0.002}_{-0.003}$&$3.2\;^{+0 .1}_{-0.2}$&$1.7\;^{+0.4}_{-0.3}$      & 48 &$3800\;^{+100}_{-200}$ & 41.6 \\
1992ba &$13\;^{+2}_{-1}$&$17\;^{+2}_{-2}$&$0.016\;^{+0.003}_{-0.002}$&$3.2\;^{+0.2}_{-0.4}$&$2.1\;^{+0.5}_{-0.7}$       & 42 &$3500\;^{+200}_{-300}$ & 41.5 \\
2003Z &$13\;^{+2}_{-1}$&$19\;^{+2}_{-2}$&$0.006\;^{+0.001}_{-0.002}$&$2.2\;^{+0.2}_{-0.1}$&$1.1\;^{+0.3}_{-0.2}$        & 28 &$4000\;^{+200}_{-200}$ & 41.25 \\
1997D &$10\;^{+0.5}_{-0.5}$&$17\;^{+3}_{-2}$&$0.008\;^{+0.001}_{-0.002}$&$2.1\;^{+0.2}_{-0.2}$&$0.9\;^{+0.3}_{-0.2}$    & 32 &$3900\;^{+200}_{-200}$ & 41.15 \\
1994N &$16\;^{+1}_{-3}$&$15\;^{+2}_{-2}$&$0.0068\;^{+0.0003}_{-0.0003}$&$2.1\;^{+0.2}_{-0.2}$&$0.8\;^{+0.3}_{-0.2}$     & 38 &$4200\;^{+300}_{-200}$ & 41.4 \\
2001dc &$10\;^{+1}_{-1}$&$12\;^{+2}_{-2}$&$0.0058\;^{+0.0005}_{-0.0007}$&$1.9\;^{+0.3}_{-0.2}$&$0.5\;^{+0.3}_{-0.1}$    & 27 &$4000\;^{+200}_{-200}$ & 41.1 \\
1999eu &$8\;^{+0.4}_{-0.6}$&$12\;^{+2}_{-1}$&$0.003\;^{+0.0005}_{-0.0004}$&$1.8\;^{+0.3}_{-0.1}$&$0.5\;^{+0.3}_{-0.1}$& 38 &$3600\;^{+200}_{-100}$ & 40.9 \\
1999br &$7\;^{+0.4}_{-0.6}$&$15\;^{+2}_{-2}$&$0.0021\;^{+0.0002}_{-0.0002}$&$1.8\;^{+0.1}_{-0.2}$&$0.6\;^{+0.1}_{-0.2}$ & 33 &$3400\;^{+100}_{-200}$ & 40.85 \\
\noalign{\smallskip}\hline
\end{tabular}

\begin{flushleft}
$R_0$ is the initial radius of the ejected envelope at the onset of
expansion

$M_{env}$ is the ejected envelope mass

$M_{Ni}$ is the ejected $^{56}$Ni mass

$V_0$ is the velocity of the homologously expanding envelope at the
outer shell

$E$ is the initial thermal+kinetic energy of the ejected envelope

$t_{rec,0}$ is the time when the envelope starts to recombine

$T_{eff}$ is the effective temperature during recombination

$L_p$ Luminosity (BVRI bands) at $t_{rec,0}$ (plateau luminosity)
\end{flushleft}
\end{table}

%
%
%
\begin{figure}
\centering
\includegraphics[height=6.0cm]{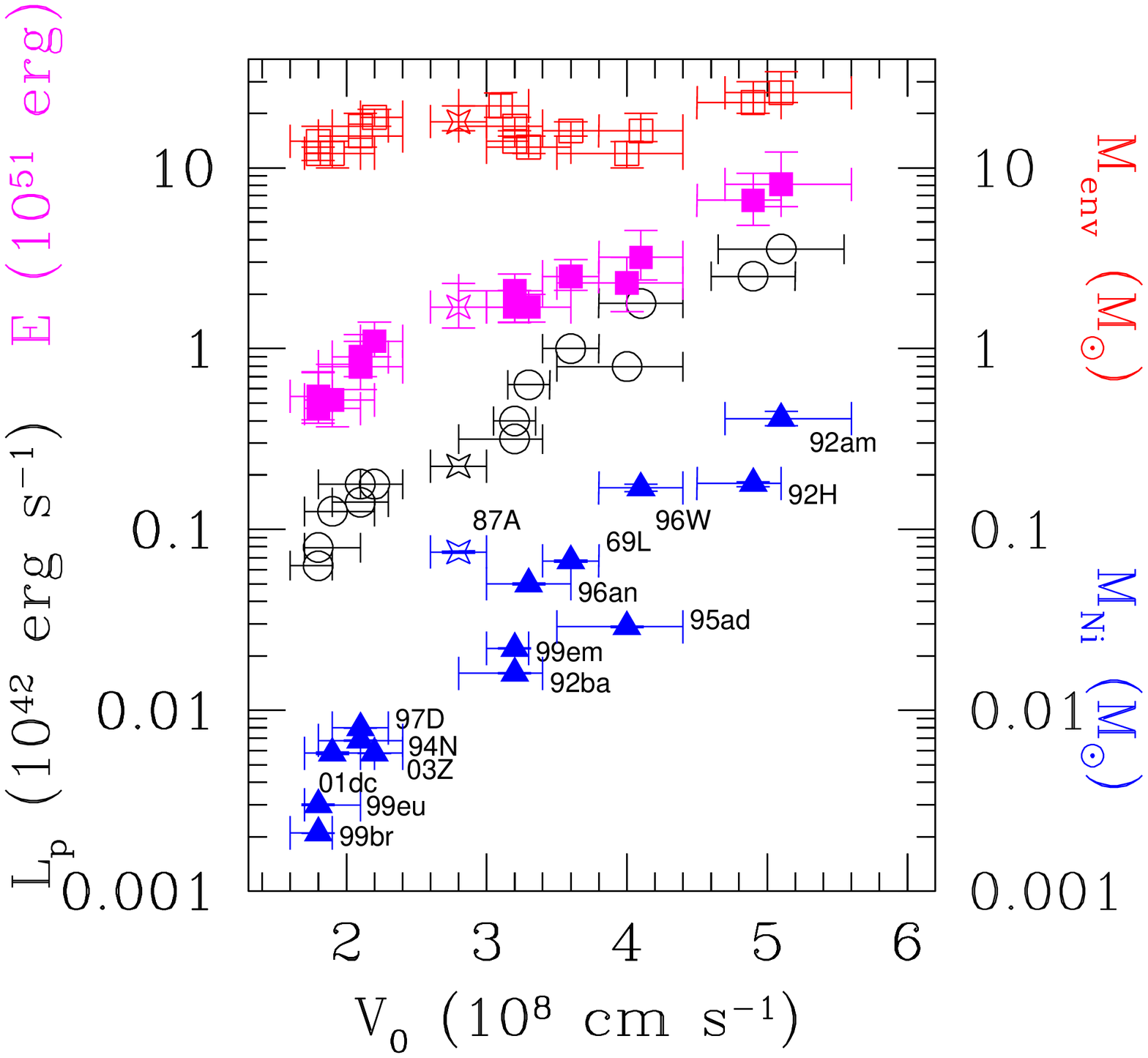}
\includegraphics[height=5.7cm]{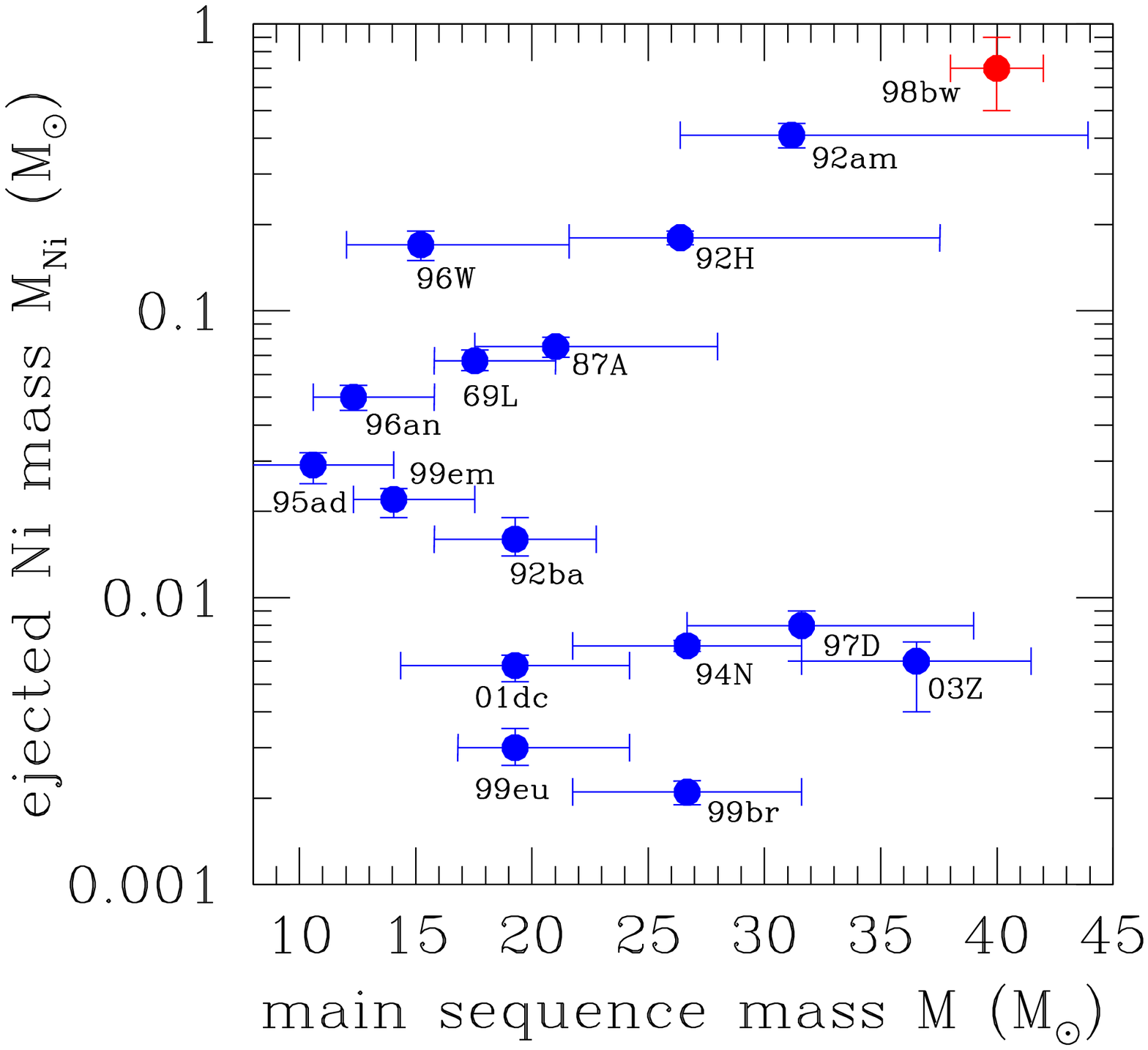}
%
%
\caption{{\it Left}: Luminosity $L_p$ ({\it circles}), energy $E$ ({\it filled squares}), envelope mass $M_{env}$ ({\it open squares}) and $^{56}$Ni mass $M_{Ni}$ ({\it triangles}) vs expansion velocity $V_0$ for the SNe of our sample. The asterisks denote SN 1987A. {\it Right}: Ejected $^{56}$Ni mass $M_{Ni}$ vs inferred progenitor main sequence mass $M$. SN 1998bw is shown for comparison.}
\label{fig:1}       
\end{figure}

As shown in Figure~\ref{fig:1} (left panel), the inferred physical
parameters of the ejected envelope are strongly correlated. All
quantities appear to vary continuously with the plateau luminosity
$L_p$ or, alternatively, with the expansion velocity of the envelope
at the outer shell $V_0$, which coincides with the photospheric
velocity measured at the onset of recombination. In particular, the
$^{56}$Ni mass increases with $V_0$ over several orders of magnitude.
The sole exception is the ejected envelope mass $M_{env}$ that, within
the estimated errors, does not show any definite tendency to vary with
the other parameters. Only at high velocities (and luminosities) does
$M_{env}$ increase slightly with $V_0$.

Correlations between the observed luminosity and photospheric velocity
at 50 days after the explosion, and between the observed luminosity
and inferred ejected $^{56}$Ni mass have recently been reported by
Hamuy \cite{hamuy03}. Our results confirm his findings and clarify
that the physical variable associated to the photospheric velocity at
50 days after the explosion is the expansion velocity $V_0$.
It is worth noting also that LL, Ni-poor SNe, such as SN 1997D and SN
1999br, do not appear to occupy a separate area of the diagram but,
instead, populate the low energy tail of the correlation, showing a
continuum variation of their parameters with respect to their more
energetic cousins.


The very weak dependence of $M_{env}$ on the other parameters has
important consequences for the nature of the progenitor and the
compact remnant.
We derived a rough estimate of the progenitor main sequence mass $M$
assuming no mass loss and a simple but physically plausible ``mixing
recipe'', that is the fraction of Carbon-Oxygen-Helium mass $M_{mix}$
mixed into the hydrogen layer and ejected increases with increasing
$^{56}$Ni yield or expansion velocity ($f_{mix}=M_{mix}/M_{env}=0.15$
for the LL events, $f_{mix}=0.4$ for the ``normal'' Type II SNe,
$f_{mix}=0.45$ for the high luminosity objects). The results are not
strongly dependent on the specific prescription for mixing, as long as
it is taken to increase with $M_{Ni}$ or $V_0$. From this, assuming no
rotation, we then estimate the hydrogen mass from
$M_H=M_{env}-M_{mix}=(1-f_{mix})M_{env}$ and the main sequence mass
from the approximate expression $M=2.9(1-f_{mix})M_{env}-10.3$ (see
e.g. \cite{arn96}). Including mass loss would result in larger values
of $M$, especially for $M\geq 20 M_\odot$, with a significant
dependence on metallicity.
We find that both high and LL SNe have massive progenitors with $M\geq
20-25 M_\odot$. This follows, for the first, from the large inferred
value of $M_{env}$ while, for the second, from the fact that $f_{mix}$
is small. Because ``normal'' and LL events have similar ejected
envelope masses $M_{env}$ but rather different mixing fractions, the
first have comparatively less massive progenitors ($12\leq M \leq 20
M_\odot$). For the ``normal'' and high luminosity SNe a large fraction
of the ejected envelope mass comes from the Carbon-Oxygen-Helium layer
that was successfully ejected, while in the LL events $M_{env}$
essentially measures the ejected Hydrogen mass. Despite the large
errors, we find that only LL SNe appear to have progenitors with
masses significantly in excess of $M_{env}$. Thus, they may have
undergone significant fallback, as suggested by Zampieri et
al. \cite{zamp03}, and harbor rather massive black holes.

In Figure~\ref{fig:1} (right panel) we show the plot of the ejected
$^{56}$Ni mass versus the inferred progenitor main sequence
mass. Again, albeit the errors are very large, it is possible to
recognize that Type II SNe populate different regions in this
diagram. In particular, as originally suggested by Iwamoto et
al. \cite{iwa00}, there appears to be a bimodal behavior above
$\approx 20 M_\odot$, with high luminosity events populating the high
$^{56}$Ni tail (close to the area occupied by hypernovae as SN 1998bw)
and LL objects filling the low $^{56}$Ni yield region. The reason for
the large spread in ejected $^{56}$Ni and the non monotonic behavior
of the $M-M_{Ni}$ relation is not clear. Perhaps, as suggested by
Maeda and Nomoto \cite{maeda03}, in luminous events a large amount of
angular momentum is retained by the post-shock envelope causing the
formation of jets and an enhanced energy release along the jet axes,
whereas ``normal'' and LL events may have more spherical shapes. It
could also be that different metallicities and mass loss histories
prior to explosion play an important role, with high luminosity events
having more powerful winds while LL ones retain almost all their
hydrogen envelope until explosion.

%
%
%
%
%

%
%



\printindex
\end{document}